\documentclass[%
 reprint,%
 amssymb, amsmath,%
 aip,cha,jmp
]{revtex4-1}
\usepackage{graphicx}
\usepackage{dcolumn}
\usepackage{caption}
\usepackage{subcaption}
\usepackage{float}
\usepackage{color}
\usepackage[normalem]{ulem}
\usepackage{booktabs}
\begin{document}


\title{Extreme ultraviolet radiation induced defects in single-layer graphene} 

\author{A. Gao}
\email{a.gao@differ.nl.}
 \affiliation{FOM-Dutch Institute for Fundamental Energy Research, Edisonbaan 14,3439 MN Nieuwegein, the Netherlands.}
 \affiliation{XUV group, MESA+ Institute for Nanotechnology, PO Box 217, University of Twente, 7500 AE, Enschede, the Netherlands.}
\author{E. Zoethout}
\affiliation{FOM-Dutch Institute for Fundamental Energy Research, Edisonbaan 14,3439 MN Nieuwegein, the Netherlands.} 
\author{J.M. Sturm}
\affiliation{FOM-Dutch Institute for Fundamental Energy Research, Edisonbaan 14,3439 MN Nieuwegein, the Netherlands.}
\affiliation{XUV group, MESA+ Institute for Nanotechnology, PO Box 217, University of Twente, 7500 AE, Enschede, the Netherlands.}
\affiliation{Materials innovation institute M2i, Mekelweg 2, 2628 CD Delft, the Netherlands.}

\author{C.J. Lee}
 \affiliation{FOM-Dutch Institute for Fundamental Energy Research, Edisonbaan 14,3439 MN Nieuwegein, the Netherlands.}
 \affiliation{XUV group, MESA+ Institute for Nanotechnology, PO Box 217, University of Twente, 7500 AE, Enschede, the Netherlands.}
\author{F. Bijkerk}
 \affiliation{FOM-Dutch Institute for Fundamental Energy Research, Edisonbaan 14,3439 MN Nieuwegein, the Netherlands.}
	\affiliation{XUV group, MESA+ Institute for Nanotechnology, PO Box 217, University of Twente, 7500 AE, Enschede, the Netherlands.}
\date{\today}

\begin{abstract}
We study extreme ultraviolet~(EUV) radiation induced defects in single-layer graphene. Two mechanisms for inducing defects in graphene were separately investigated: photon induced chemical reactions between graphene and background residual gases, and breaking sp$^2$ bonds, due to photon and/or photoelectrons induced bond cleaving. Raman spectroscopy shows that D peak intensities grow after EUV irradiation with increasing water partial pressure in the exposure chamber. Temperature-programmed desorption~(TPD) experiments prove that EUV radiation results in water dissociation on the graphene surface. The oxidation of graphene, caused by water dissociation, is triggered by photon and/or photoelectron induced dissociation of water. Our studies show that the EUV photons cleave the sp$^2$ bonds, forming sp$^3$ bonds, leading to defects in graphene.
\end{abstract}

\pacs{61.48.De}

\maketitle 
\thispagestyle{plain}
\pagestyle{plain}

\section{\label{sec:intro}Introduction}

Graphene, a two-dimensional hexagonal packed sheet of carbon atoms, has attracted a lot of attention from different research fields due to its unique physical and chemical properties~\cite{geim2007rise,geim2009graphene,novoselov2005two,zhang2005experimental,han2007energy,bolotin2008ultrahigh,lee2008measurement,bunch2007electromechanical}. Graphene can act as a diffusion barrier by providing physical separation between an underlying substrate and reactant gases. Indeed, studies show that graphene is highly impermeable to gases~\cite{bunch2008impermeable}. Furthermore, single-layer graphene is also highly transparent~\cite{nair2008fine}, which makes it a promising protection layer for optical devices, such as mirrors, lenses and screens. Additionally, graphene can be grown on large scales by chemical vapor deposition~(CVD) and transferred to arbitrary substrate~\cite{reina2008large}, making the range of potential applications very wide. 

Graphene devices may be used in harsh environments e.g., space applications, soft x-ray systems and extreme ultraviolet~(EUV) lithography systems. Since the light source of current EUV lithography systems operates at a wavelength of 13.5~nm, the optical system is in vacuum, but with a significant partial pressure of water. The physical and chemical stability of graphene in such an environment is of critical importance if it is to be used as a part of an optical component. Early experimental work showed that defects were generated in multilayer graphene after EUV exposure. And, even in a reducing environment, a small partial pressure of oxidizing agents~(water) may cause oxidation~\cite{gao2013extreme}. The damage to graphene observed in our previous study was attributed to two possible effects: photo-induced chemistry between graphene and background residual gases, and breaking sp$^2$ bonds, due to EUV photon and/or photoelectron induced bond cleaving.
 
In this work, we investigate both of these possible EUV-induced damage mechanisms on CVD grown single layer graphene. In previous work~\cite{gao2013extreme}, the oxidative damage to graphene was tentatively attributed to the presence of a reactive low density water plasma. But in the ref~\cite{gao2013extreme}, the amount of water~(which was the main source of oxidative damage to graphene) adsorbed to the graphene surface was not controlled, nor was it possible to perform experiments without a substantial plasma-surface interaction. In this work, the influence of the EUV-induced water plasma is controlled in two different ways. In the first set of experiments, the partial pressure of water in the vacuum chamber was set by introducing water vapor into the interaction chamber. In the second set of experiments, water layers were deliberately deposited on a cold graphene surface~(83~K). In the first set of experiments, the graphene is exposed to an EUV-induced plasma, which contains different concentrations of water plasma. In the second set of experiments, the solid water layer can be ionized into much denser plasma under EUV irradiation. In addition, since this water layer is physisorbed onto the graphene surface, the reaction probability is much higher compared with that in the first set of experiments. By comparing the nature and density of defects induced in these two experiments, it is possible to determine the relative contribution of the EUV-induced plasma to the damage observed in graphene after exposure. 

Finally, in a third set of experiments, naturally accumulated hydrocarbon contamination~(0.7~nm) on a graphene surface was used as a barrier layer between the residual water and graphene surface. In this way, it was possible to study the damage to graphene, while minimizing the reaction rate between graphene and water plasma.

In our experiments, the graphene was characterized using Raman spectroscopy, X-Ray photoelectron spectroscopy~(XPS), and scanning electron microscopy~(SEM), before and after exposure. The ratio of the D and G Raman spectral features is used as a measure of how well-ordered the graphene crystalline structure is, while XPS is used to determine the amount of oxidation. SEM was used to qualitatively compare the graphene layer completeness. The adsorbed water was examined using temperature-programmed desorption~(TPD) spectroscopy to understand the water layer morphology, water-water, and water-graphene interactions

\section{\label{sec:exp}Experiments}
For the first and second sets of experiments, single layer graphene samples, obtained from Graphene Supermarket Inc., were grown on copper substrates by CVD and transferred to SiO$_2$/Si substrates. The substrate size was 10~mm~x~10~mm with a 285~nm thick layer of SiO$_2$. For the third set of experiments, single layer graphene samples on Cu substrates were purchased from Graphenea, which were stored in ambient condition for several months, leading to a layer of hydrocarbons about 0.7~nm thick on top of the graphene layer. For the first and third sets of experiments, the graphene samples were exposed to EUV from a Xe plasma discharge source~(Philips EUV Alpha Source 2) with a repetition rate of 500 Hz. The EUV beam profile has a Gaussian distribution with FWHM= 3~mm. The EUV intensity at the sample surface was estimated to be $5~W/cm^2$ with a dose of $10~mJ/cm^2$ per pulse. The base pressure of the exposure chamber was 2x10$^{-9}$~mbar. For the second set of experiments, the graphene samples were exposed with EUV intensity of $0.05~W/cm^2$, and the base pressure of the exposure chamber was 1x10$^{-9}$~mbar.

All exposed samples were characterized using Raman spectroscopy, XPS, and SEM after exposure. Raman spectra were collected with a home-built system, based on a 532~nm solid state laser system and a Solar M266 spectrometer with a resolution of 1~cm$^{-1}$. The illumination intensity was set at $200~W/cm^2$. The collection optics and pixel size of the detector result in a spatial resolution of $100\times100~um^2$. The collection efficiency of the detector system was calibrated using the HG-1 Mercury Argon Calibration Light Source and AvaLight-D(H)-S Deuterium-Halogen Light Source. XPS was measured by using a monochromatic Al-K¦Á, Thermo Fisher Theta probe with a beam footprint of 1~mm diameter.

\begin{figure*}[!htb]
        \begin{subfigure}{0.4\textwidth}
                \includegraphics[width=\textwidth]{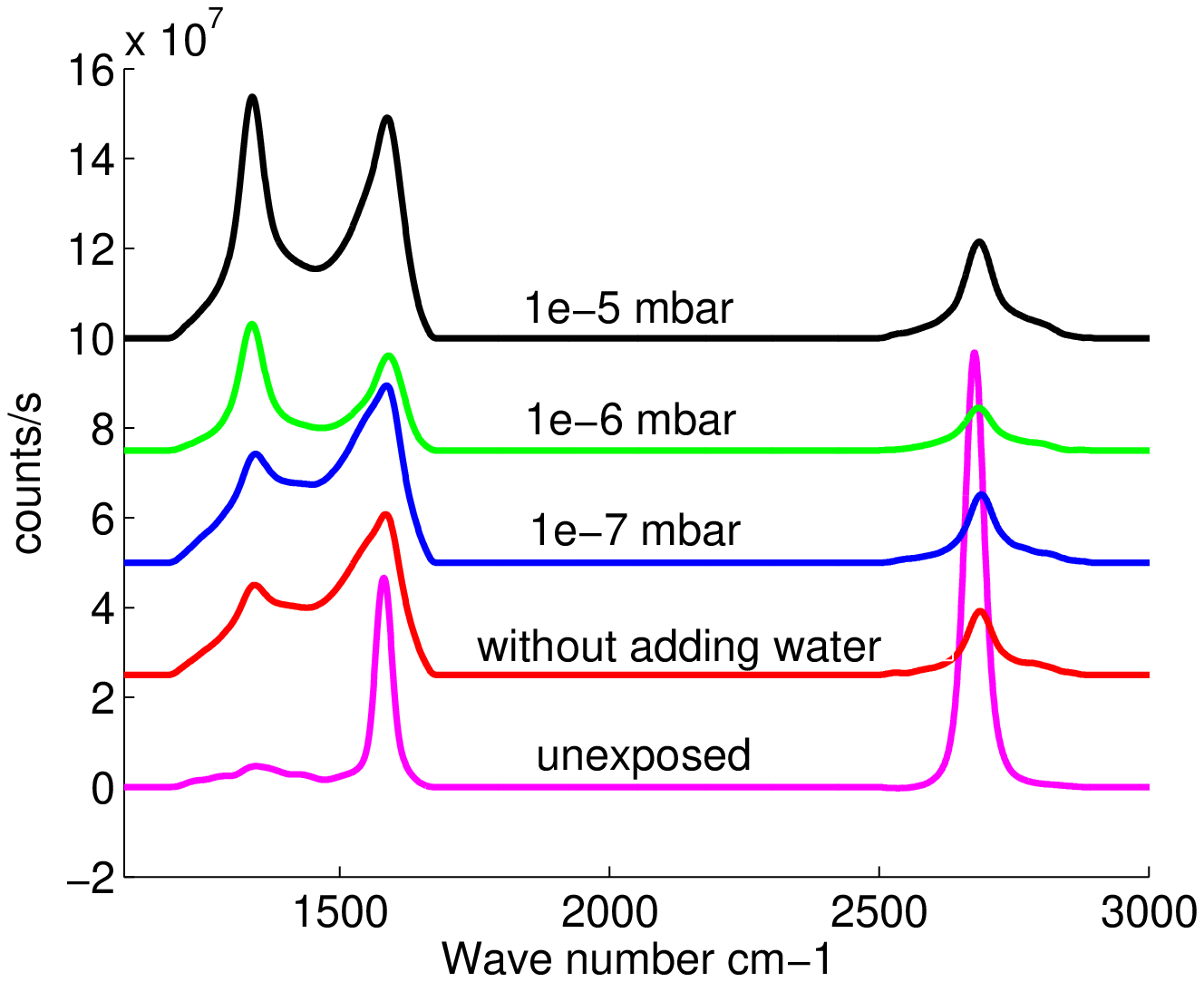}
                \caption{}
               \label{fig:1a}                
        \end{subfigure}~
        \begin{subfigure}{0.4\textwidth}
                \includegraphics[width=\textwidth]{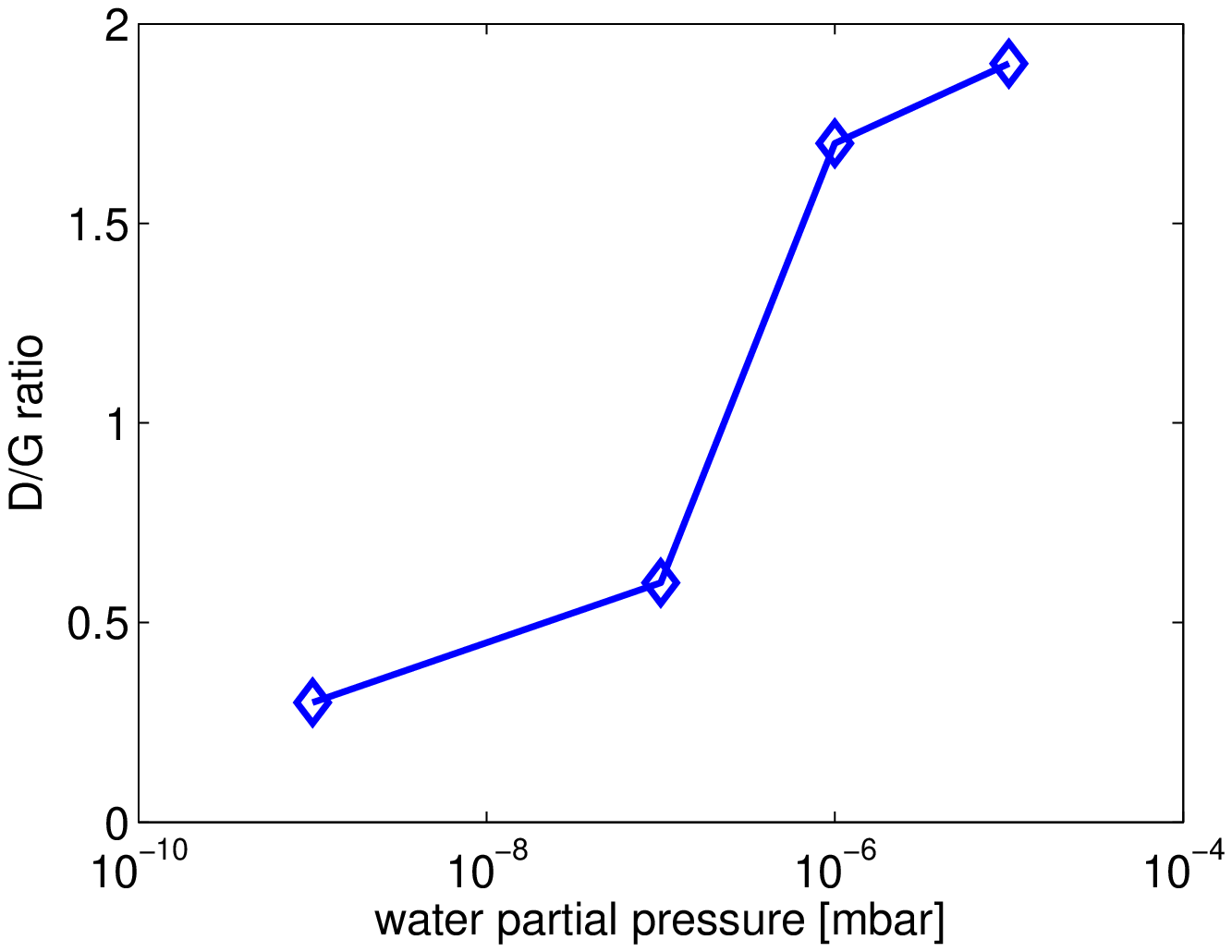}
                \caption{}
                \label{fig:1b}
        \end{subfigure}%
        \caption{(color online) (a) The Raman spectra of the graphene samples on SiO$_2$/Si substrate exposed to EUV under different water partial pressure. (b) The \textit{I(D)/I(G)} ratio as function of the water partial pressure. The x axis is in log scale.}\label{fig:figure1}
\end{figure*}

\section{\label{sec:results}Results and Discussion}
\subsection{Photo-induced plasma of the residual water}

Four graphene samples~(single layer graphene on SiO$_2$/Si substrate) were exposed to EUV for 30~min with various water partial pressures. One of the graphene samples was exposed without adding water into the chamber (water partial pressure less than 10$^{-9}$~mbar), while another three samples were exposed with water partial pressures of 1x10$^{-7}$~mbar, 1x10$^{-6}$~mbar, and 1x10$^{-5}$~mbar, respectively. These four samples were kept at 289~K by backside cooling during the exposure. After exposure, all the samples were examined by Raman spectroscopy and XPS. A typical Raman spectrum of graphene has three prominent features i.e., D, G and 2D peaks, located at approximately 1350~cm$^{-1}$, 1580~cm$^{-1}$, and 2680~cm$^{-1}$. The G peak is a first order Raman scattering process, corresponding to an in plane stretching of sp$^2$ bonds. The D band is due to the breathing modes of six-atom rings, and requires a defect for activation. The ratio of \textit{I(D)/I(G)} is commonly used to quantify the defect density~\cite{ferrari2013raman}. Fig.~\ref{fig:1a} shows the Raman spectra of the reference graphene sample~(unexposed) and the exposed graphene samples. It clearly shows that the D peak height grows after exposure for all samples, indicating defect generation. Furthermore, the dependence of the \textit{I(D)/I(G)} ratio on water partial pressure is clearly indicated in Fig.~\ref{fig:1b}.

\begin{figure*}[!htb]
        \begin{subfigure}{0.4\textwidth}
                \centering
                \includegraphics[width=\textwidth]{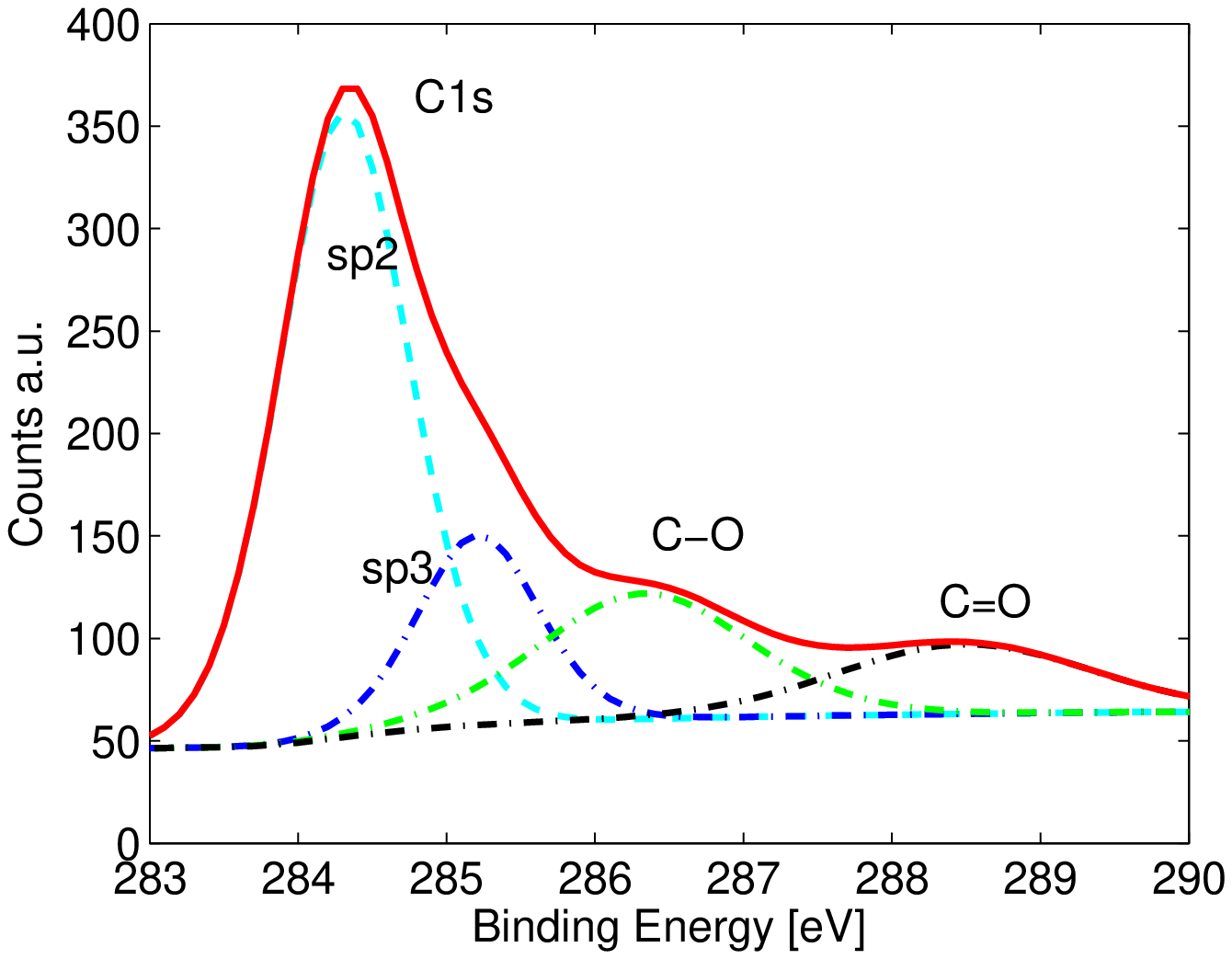}
                \caption{}
               \label{fig:2a}                
        \end{subfigure}~
        \begin{subfigure}{0.4\textwidth}
                \centering
                \includegraphics[width=\textwidth]{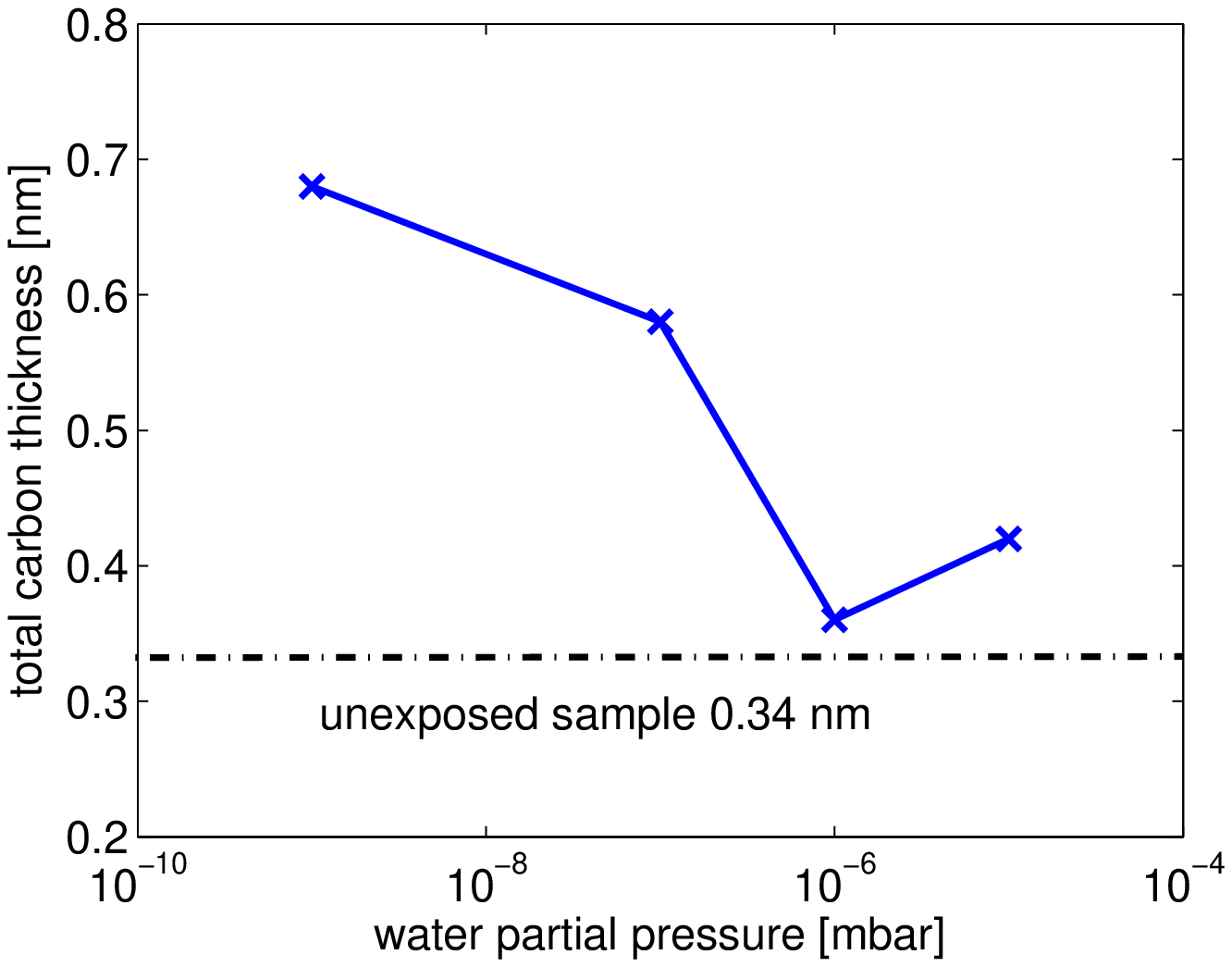}
                \caption{}
                \label{fig:2b}
        \end{subfigure}%
        \caption{(color online) (a) The curve fitting results of the XPS spectrum for the graphene samples on SiO$_2$/Si substrate exposed to EUV with water partial pressure of 1x10$^{-5}$ mbar. The solid curve indicates the C1s peak while the four dot-dash curves are fitted curves. (b) The total carbon thickness of the exposed samples with various water partial pressures. Note that the unexposed sample has slightly hydrocarbon contamination on the surface due to transport and storage in atmospheric conditions. The thickness is calculated based on the angle resolved XPS measurements assuming the carbon density of 2.1~g/cm$^3$. The x axis is in log scale.}\label{fig:figure2}
\end{figure*}

The Raman spectra, however, do not clearly show the source of the increased defect density. Fig.~\ref{fig:2a} shows the curve fit for the XPS spectrum of the C1s peak of the pristine sample. The four components of the C1s spectrum, corresponding to sp$^2$ bonds in graphitic like carbon~(284.3~eV), sp$^3$ hybridization~(285.2~eV), hydroxyl~(C-O) groups~(286.1~eV), and carboxyl~(C=O) groups~(288~eV), are plotted separately for the exposed sample. The appearance of sp$^3$, C-O, and C=O bonds can be attributed to photo-induced bond cleaving and photo-induced oxidation of graphene, due to presence of water. However, Fig.~\ref{fig:2b} also shows that the total carbon thickness of the exposed samples increases, due to the well-known effect of hydrocarbon deposition during EUV exposure~\cite{madey2006surface}. The contribution of the deposited hydrocarbons to the XPS spectra makes a single interpretation of the data impossible, since the sp$^3$, C-O, and C=O contributions can also come from the hydrocarbon contamination. Note that the EUV transmission will decrease by less than 1~\% due to 0.3~nm carbon contamination~\cite{windt1998imd}, thus, the difference of EUV dose on graphene samples is negligible, and cannot explain the differences between Raman spectra for the five samples.

\subsection{Photo-induced plasma of the adsorbed water}

 \begin{figure*}[!htb]
        \begin{subfigure}{0.4\textwidth}
                \centering
                \includegraphics[width=\textwidth]{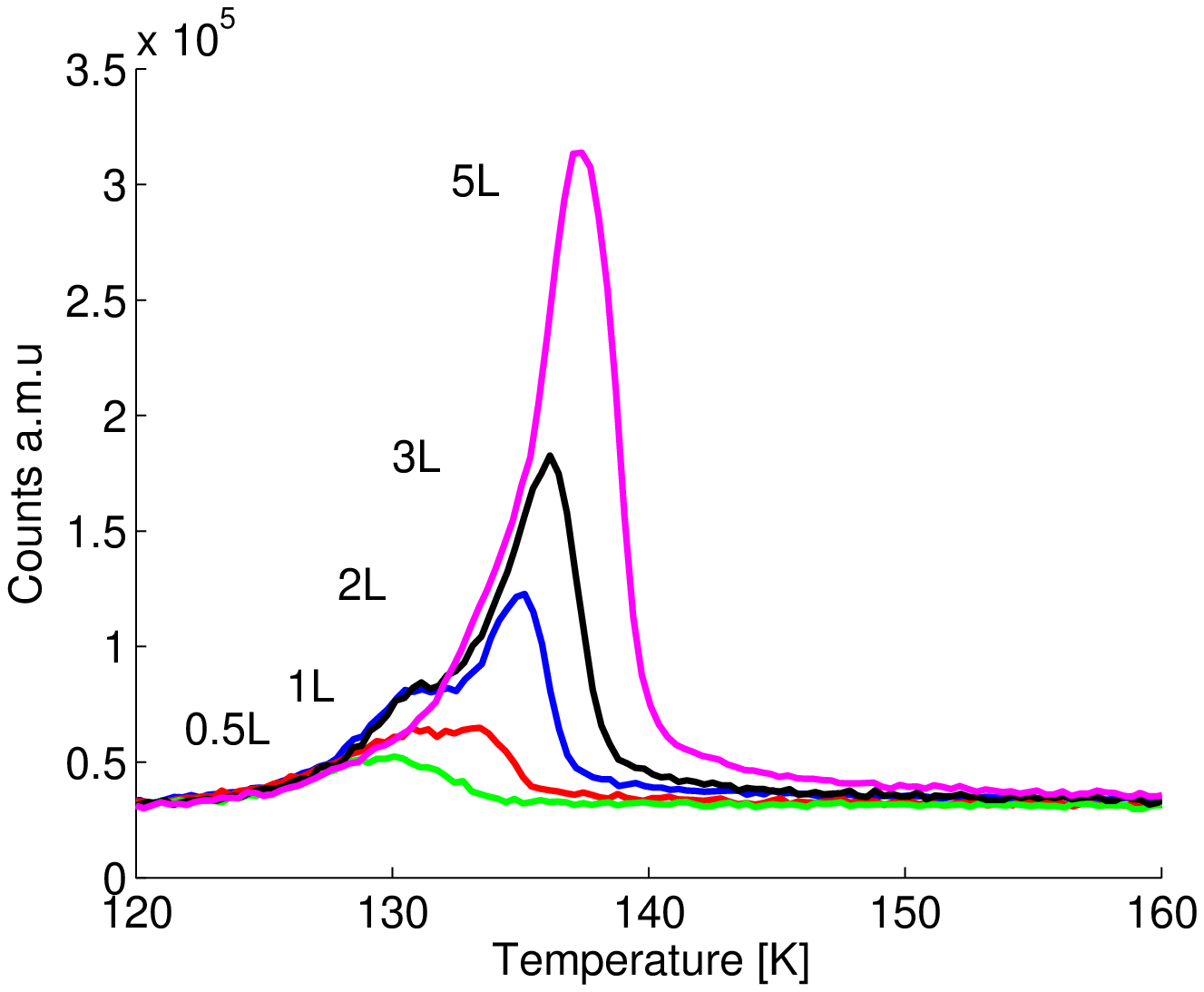}
                \caption{}
               \label{fig:3a}                
        \end{subfigure}~
        \begin{subfigure}{0.4\textwidth}
                \centering
                \includegraphics[width=\textwidth]{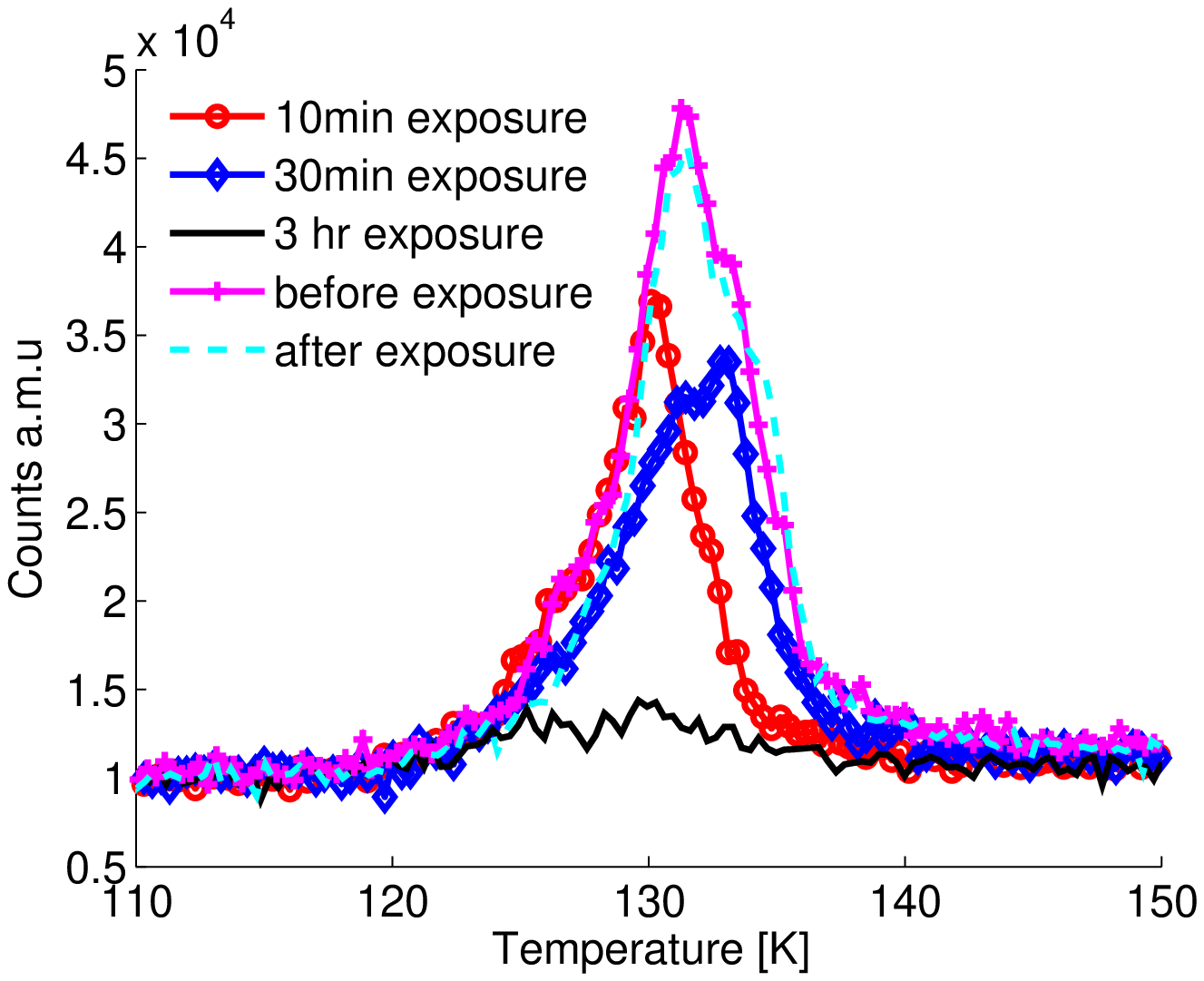}
                \caption{}
                \label{fig:3b}
        \end{subfigure}%
        \caption{(color on line) Temperature-programmed desorption spectra for H$_2$O desorption on the graphene surface (a) TPD spectra for various H$_2$O doses; (b) TPD spectra of the graphene sample under different EUV exposure time with water dosing time of 1~L. The heating rate is 0.5~K/s. The curves are manually offset to the have the same background.}\label{fig:figure3}
\end{figure*}
\begin{figure*}[!htb]
        \begin{subfigure}{0.4\textwidth}
                \centering
                \includegraphics[width=\textwidth]{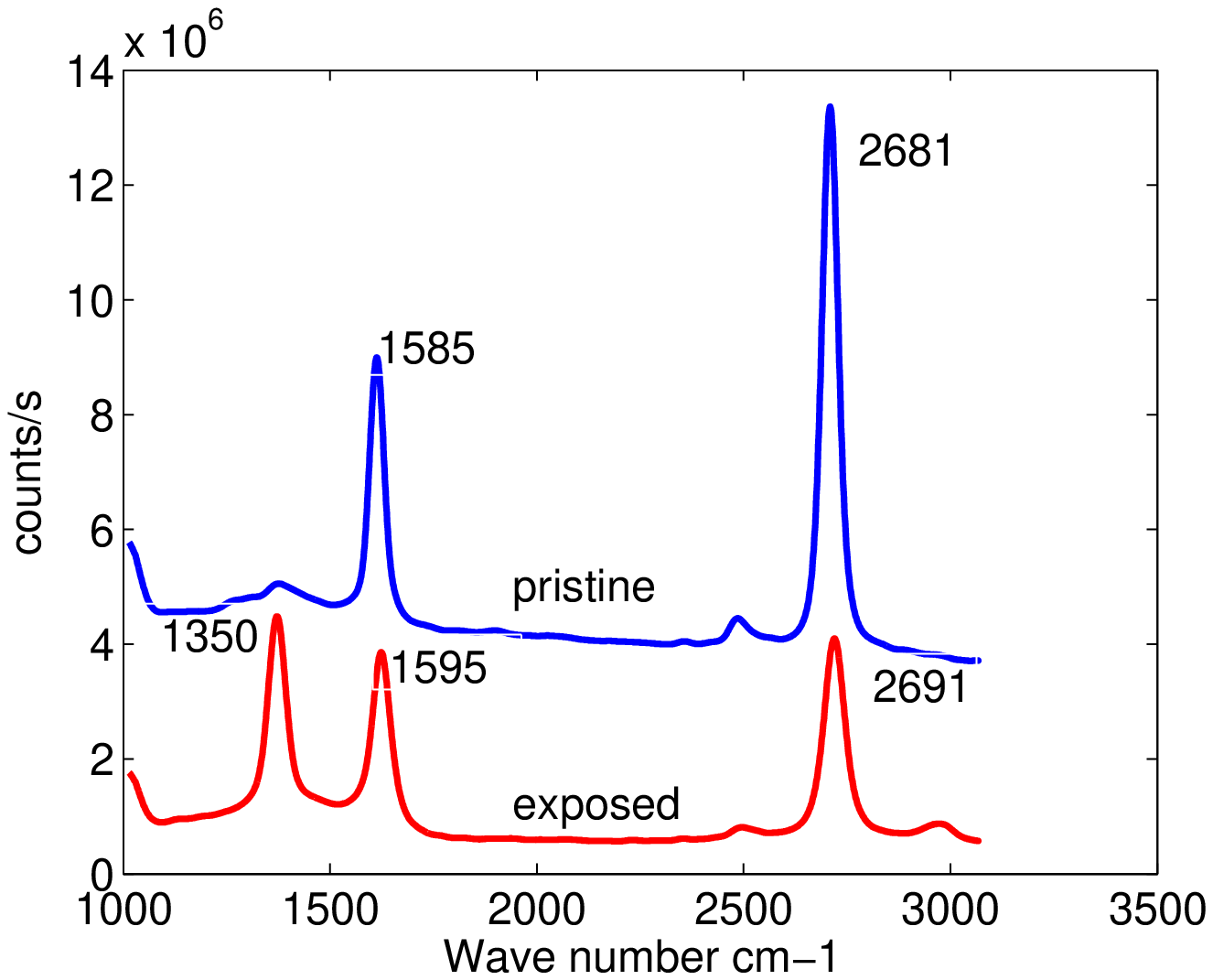}
                \caption{}
               \label{fig:4a}                
        \end{subfigure}~
        \begin{subfigure}{0.4\textwidth}
                \centering
                \includegraphics[width=\textwidth]{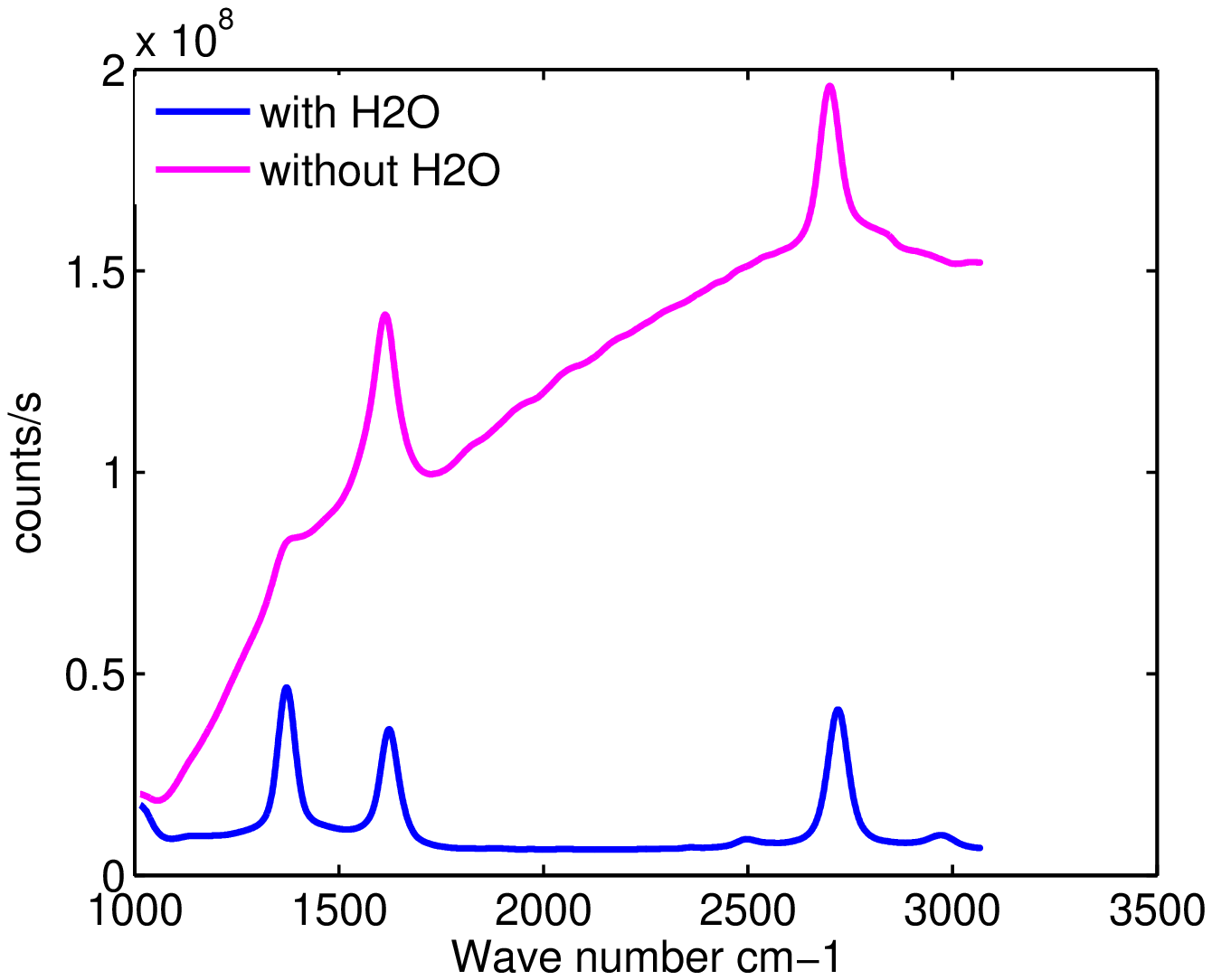}
                \caption{}
                \label{fig:4b}
        \end{subfigure}%
        \caption{(color online) (a) Raman spectra of the graphene sample before~(pristine) and after EUV radiation with water adsorbed onto the sample~(exposed).(b) Raman spectra of the samples exposed with or without water adsorbed onto graphene surface.}\label{fig:figure4}
\end{figure*}
        
In order to prevent hydrocarbon contamination, a layer of water was deliberately deposited on the graphene surface by dosing the surface with water while it was held at a temperature of 83~K. Directly before water dosing, the graphene sample was heated to 600~K to remove surface contamination, after which a calibrated surface coverage of water was adsorbed onto the sample. The water dose is expressed in Langmuir, with 1~L = 1.33x10$^{-6}$~mbar.

Temperature-programmed desorption~(TPD) spectroscopy was performed to investigate how water adsorbs on to the graphene surface. The graphene sample was heated at a rate of 0.5~K/s and water desorption was detected by a quadruple mass spectrometer. To ensure that the signal is dominated by sample surface desorption, the entrance to the mass spectrometer consists of a cone that faces the sample surface. The cone has an entrance aperture of 4~mm, located approximately 2~mm from the graphene surface. The temperature was measured with a thermocouple pressed onto the sample with a molybdenum clamp. Since this clamp has a relatively good thermal contact with the cold finger of the manipulator, the measured temperature will underestimate the actual sample temperature.

In our experimental procedure, the graphene sample was dosed with water, and TPD spectroscopy was performed prior to EUV exposure. The sample was then re-dosed with water and exposed to EUV. A second TPD spectrum was obtained from the exposed water-graphene sample. Finally, the sample was dosed with water a third time and a third TPD was performed on the exposed graphene. This procedure was repeated three times on the same sample after EUV exposure of 10~min, 30~min, and 180~min respectively. The EUV intensity in this exposure was estimated to be $100~\mu~J/cm^2$, about 100 times lower than used in the first set of experiments.

The TPD results are shown in Fig.~\ref{fig:3a}. The spectra at low water dosages and high water dosages are markedly different, indicating that the structure of the adsorbed water is different, or evolves differently as function of both temperature and initial dose. As graphene~(with certain hydrocarbon contamination) has a hydrophobic surface, water films tend to dewet upon heating. At low coverage, the water molecules form two-dimensional~(2D) clusters. Water molecules at edges of these clusters have low coordination to other water molecules, such that their desorption energy is low, resulting in the desorption peak around 130~K~\cite{souda2012nanoconfinement,bolina2005reflection,chakarov1998photoinduced}.

However, as the coverage increases, water molecules can form 3D clusters of crystalline-ice like structure, resulting in a second peak at 134~K. The position (temperature) of this peak increases with coverage, while the leading edge remains similar. This is characteristics for zero-order desorption of water molecules from ice~\cite{bolina2005reflection,chakarov1998photoinduced}. The rearrangement of the water suppresses the peak at 130~K for water doses above 3~L. At a water dose of 5~L the 130~K peak disappears, which indicates that all water molecules form multilayers or 3D ice clusters. The TPD data from Fig.~\ref{fig:3a} also indicates that water does not dissociate on adsorption to the graphene surface, thus, any water dissociation and resulting changes to the graphene are driven by EUV-induced processes as discussed in the following.

In Fig.~\ref{fig:3b}, the TPD spectra of water on graphene before and after exposure are shown for a water dose of 1~L. It is shown in Fig.~\ref{fig:3b} that the spectra of the graphene sample after exposure exhibits lower peak intensities than that of before exposure. This indicates that the amount of water on the surface decreases due to EUV-induced desorption of water. Note that, aside from the decrease in the total amount of water on the graphene surface, a second TPD peak appears at higher temperature after 30~min of exposure. This second peak is likely due to the water layer rearranging itself during EUV exposure.

EUV photons with an energy of 92~eV can also cause water dissociation by direct photon excitation or by an indirect process involving secondary electrons emitted from the substrate. Many studies have shown that the indirect process, induced by secondary electrons, dominates over the direct photon excitation in the surface photochemistry~\cite{madey2006surface,andersson2004water,zhou1991photochemistry}. Therefore, direct EUV-dissociation can be neglected and the water molecules can be dissociated according to the following reactions~\cite{madey2006surface}:
\begin{equation}
\label{eq1}
e^{-} + H_{2}O \rightarrow H^{+} + OH^{-}
\end{equation}
\begin{equation}
\label{eq2}
e^{-} + OH^{-} \rightarrow H^{+} + O^{-}
\end{equation}
Upon dissociation, the reactive species can either desorb from the surface or react (mainly atomic oxygen) with the graphene. 

Fig.~\ref{fig:4a} shows the Raman spectra of the graphene sample before and after EUV exposure with adsorbed water layer. Comparing these two spectra in Fig.~\ref{fig:4a}, there is a clear D peak intensity increase at around 1350~cm$^{-1}$, which is attributed to the presence of defects in graphene. Previous work has shown that defects in graphene can be induced by EUV radiation~\cite{gao2013extreme}, however, the shift of the G peak from 1585~cm$^{-1}$ to 1595~cm$^{-1}$ and of the 2D peak 2681~cm$^{-1}$ to 2691~cm$^{-1}$, are usually interpreted as evidence of oxide doping~\cite{mitoma2013photo}.

Fig.~\ref{fig:4b} shows the Raman spectra of samples exposed with and without water adsorbed on to the graphene surface. The two spectra differ in both the \textit{I(D)/I(G)} ratio and the spectrum fluorescence background. Clearly, the \textit{I(D)/I(G)} ratio of the sample exposed with water is much higher than that of the sample exposed without water, indicating more defects were generated in the former sample. The water layer adsorbed to the graphene surface will result in a very dense water plasma, and such a plasma should have a faster reaction rate than that of water plasma generated in the background residual gas. The large fluorescence signal in the sample exposed without water is attributed to hydrocarbon contamination.

\begin{figure}[!htb]
                \centering
                \includegraphics[width=0.4\textwidth]{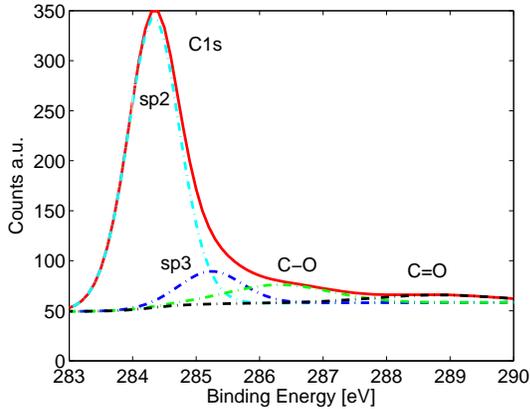}
            \caption{(color online) The XPS spectra of the graphene sample exposed to EUV with adsorbed water onto graphene. The solid curve indicates the C1s peak while the four dot-dash curves are fitted curves.}\label{fig:figure5}
\end{figure}
        
\begin{figure*}[!htb]
        \begin{subfigure}{0.4\textwidth}
                \centering
                \includegraphics[width=\textwidth]{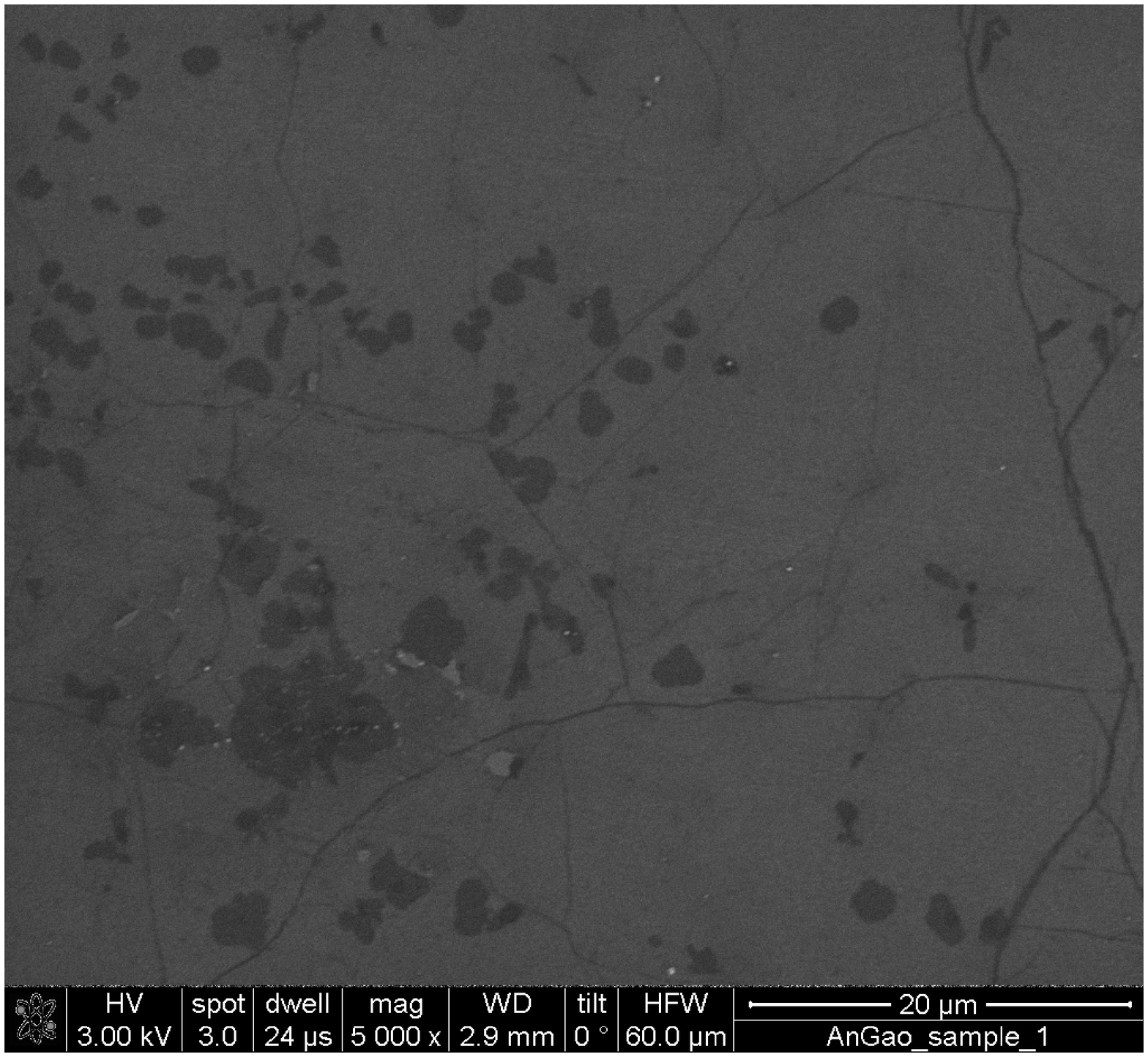}
                \caption{unexposed sample}
               \label{fig:6a}                
        \end{subfigure}~
        \begin{subfigure}{0.4\textwidth}
                \centering
                \includegraphics[width=\textwidth]{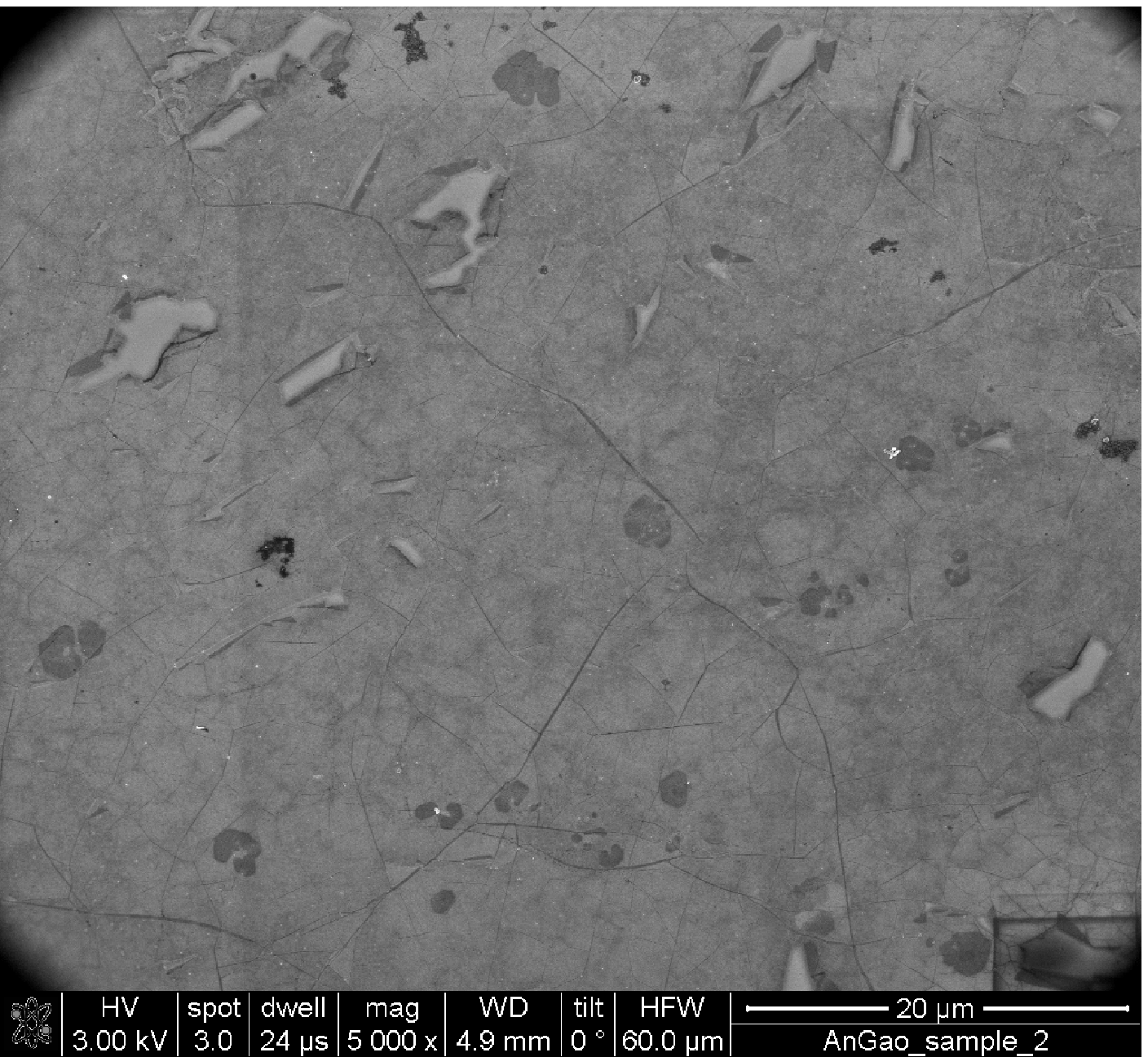}
                \caption{exposed sample}
                \label{fig:6b}
        \end{subfigure}%
        \caption{SEM images of the pristine sample, and the sample exposed to EUV with adsorbed water showing PMMA residue, cracks and holes. The PMMA appears as the darker patches in both images.}\label{fig:figure6}
\end{figure*}

Fig.~\ref{fig:figure5} shows the curve fit for the C1s peak of the sample exposed to EUV with adsorbed water. The components of the C1s spectrum: sp$^2$, sp$^3$, C-O, and C=O, are plotted. The atomic concentration of each component and full width half maximum of the spectral components for the pristine sample, and samples exposed to EUV with and without adsorbed water are summarized in Tab.~\ref{tab:table1}.  The broadening of the sp$^2$ peak from 0.88~eV to 0.98~eV is usually evidence of a transition from a highly ordered graphite-like carbon to a less ordered carbon state, supporting the conclusion from the Raman results that the exposed graphene has more defects. The changes in C-O and C=O concentrations are not significant, considering that the uncertainty in the atomic concentration is $\pm2.5~at.\%$. The C-O and C=O bonds are most likely to be due to residual poly~(methyl meth acrylate), used during the transfer process, and clearly observed on all samples using SEM (see Fig.~\ref{fig:figure6}).

From the SEM images in Fig.~\ref{fig:figure6}, it is apparent that there is less PMMA on the graphene after exposure. Cracks and holes are also observed in the SEM images for the exposed samples, indicating an oxidative etching effect.  We can see that the exposed graphene samples show more cracks and holes than the pristine sample. And these holes and cracks are predominantly along graphene grain boundaries, a phenomenon also observed for thermal oxidation of graphene~\cite{liu2008graphene,surwade2012thermal}. The formation of C-O or C=O groups as a result of oxidation is due to fully dissociated water~(e.g., atomic oxygen), or partially dissociated water~(e.g., OH groups). The small holes in the pristine sample may come from the transfer process.

\begin{table*}
\caption{\label{tab:table1}Atomic concentration of C1s component, the total carbon thickness, and FWHM of the pristine sample, and samples exposed to EUV with and without adsorbed water. The total thickness is calculated based on the angle resolved XPS measurements assuming the carbon density of 2.1~g/cm$^3$.} 
\begin{ruledtabular}
\begin{tabular}{lcccccc}
Sample&C $sp^2$~(at.\%)&C $sp^3$~(at.\%)&C-O~(at.\%)&C=O~(at.\%)&$sp^2$ FWHM~(eV)&Total thickness~(nm)\\
 \hline
Pristine~(unexposed)&20.4&4.4&2.4&1.0&0.88&0.34\\
Exposed with adsorbed H$_2$O&19.3&2.7&2.4&1.1&0.98&0.30\\
Exposed without adsorbed H$_2$O&31.5&6.4&6.4&3.7&1.03&0.74\\
\end{tabular}
\end{ruledtabular}
\end{table*}

It is also noted that the total thickness of carbon in the exposed sample decreases, due to both the etching of graphene and the removal of PMMA. This can be seen in the XPS data in Tab.~\ref{tab:table1}, where the concentration of the sp$^2$ and sp$^3$ components decrease, while both the C-O and C=O contributions remain almost unchanged. Cleaved sp$^2$ bonds can either form sp$^3$ bonds, or through oxidation form both sp$^3$ bonds and C-O or C=O bonds. The combination of decreasing carbon, but increasing oxidation is indicative that the PMMA is preferentially removed. However, the unchanged sp$^3$ and oxygen content~(despite the disruption to the graphene and the overall removal of carbon) indicates that two competing processes are present. The removal of PMMA is compensated for by the oxidation of the graphene. 

Until now, we have shown that from Raman results, \textit{I(D)/I(G)} ratio grows as the water partial pressure increases, indicating more defects are forming in graphene. Together with the XPS data, it suggests that oxidation, triggered by the EUV radiation in the presence of water, results in the graphene being etched, forming cracks and holes, which is similar to the effect observed during thermal oxidation of graphene. In the case of EUV exposure, the oxidation process originates from the dissociation of water by EUV. However, to determine the relative contributions of EUV-induced oxidation and EUV-induced bond cleaving, the EUV-induced plasma must be separated from the graphene surface.

\subsection{Exposing graphene with hydrocarbon contamination}

Samples of single layer graphene on Cu substrate with a hydrocarbon contamination layer were prepared. The hydrocarbon layer is used as barrier layer between the residual water and graphene surface. In this way, it was possible to study the damage to graphene, while minimizing the reaction rate between graphene and water plasma. Four graphene samples were exposed to EUV for 30 min with background gas conditions that should be either reducing or oxidizing. One of the graphene samples was exposed without modifying the background gas (10$^{-9}$~mbar, mostly water), a second sample was exposed in a water partial pressure of 1x10$^{-5}$~mbar, a third was exposed in a hydrogen partial pressure of 1x10$^{-5}$~mbar, and the last was exposed in an oxygen partial pressure of 1x10$^{-5}$~mbar. All samples were kept at 289~K by backside cooling during the exposure.  

\begin{figure*}[!htb]
        \begin{subfigure}{0.4\textwidth}
                \centering
                \includegraphics[width=\textwidth]{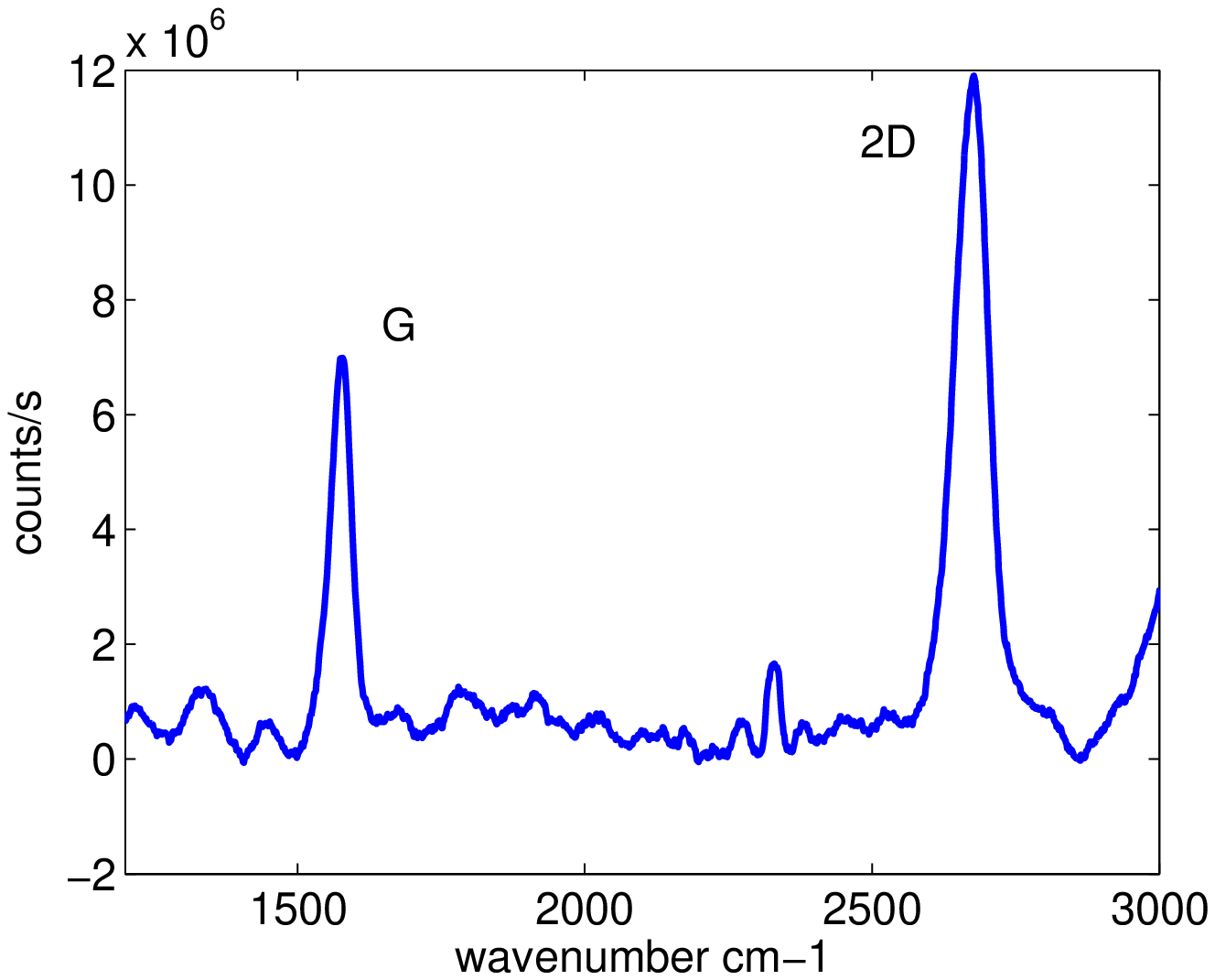}
                \caption{}
               \label{fig:7a}                
        \end{subfigure}~
        \begin{subfigure}{0.4\textwidth}
                \centering
                \includegraphics[width=\textwidth]{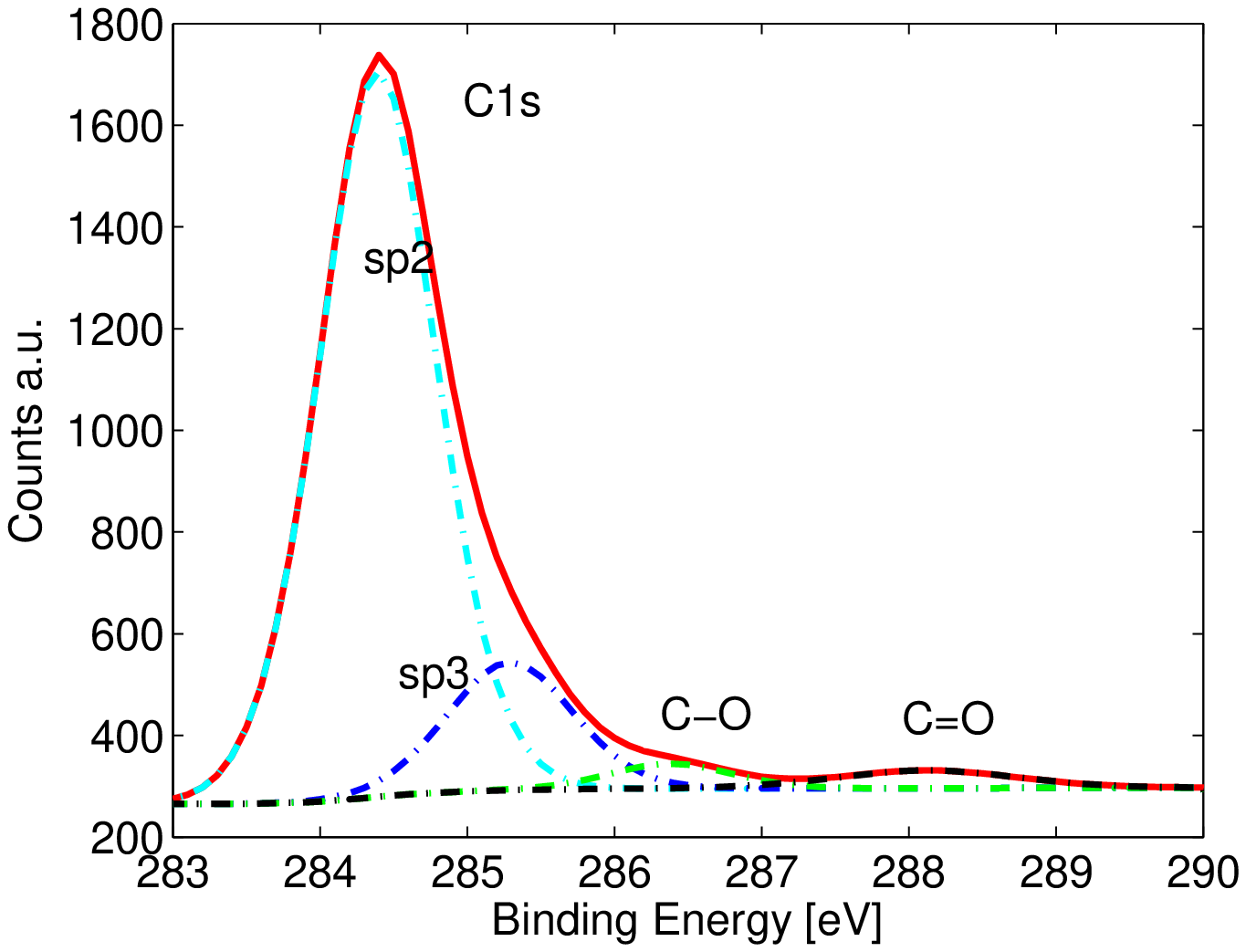}
                \caption{}
                \label{fig:7b}
        \end{subfigure}%
        \caption{(color online) Raman spectrum after fluorescence background subtraction (a) and XPS results (b) for the reference sample of monolayer graphene on Cu substrate with naturally accumulated hydrocarbon contamination.}\label{fig:figure7}
\end{figure*}

Fig.~\ref{fig:figure7} shows the Raman spectrum and XPS results for an unexposed sample. From the Raman spectra, we can see that the single layer graphene is still visible after fluorescence background subtraction, despite the presence of hydrocarbon contamination. The XPS results clearly show that, in addition to the sp2 contribution, there is a significant amount of sp$^3$, C-O, and C=O, which is attributed to the presence of hydrocarbon. Assuming a density of 2.1~g/cm$^3$, the hydrocarbon contamination layer was found to be about 0.7~nm thick from XPS measurement (see Tab.~\ref{tab:table2}). This hydrocarbon layer acts as a barrier layer between the graphene and the background residual gases in the exposure chamber. In this way, reactions between the EUV-induced plasma and the graphene can be excluded.

\begin{figure*}[!htb]
        \begin{subfigure}{0.4\textwidth}
                \centering
                \includegraphics[width=\textwidth]{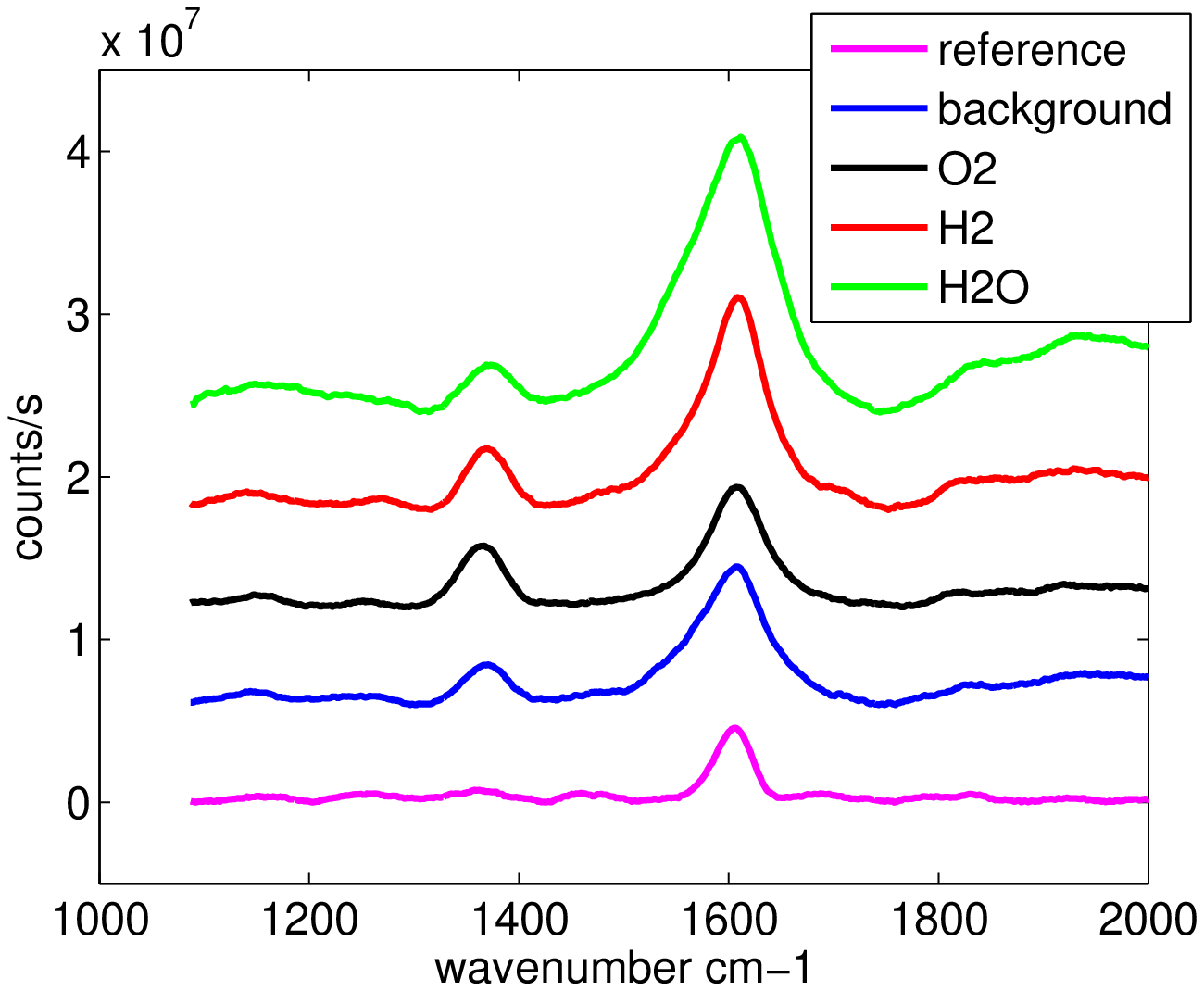}
                \caption{}
               \label{fig:8a}                
        \end{subfigure}~
        \begin{subfigure}{0.4\textwidth}
                \centering
                \includegraphics[width=\textwidth]{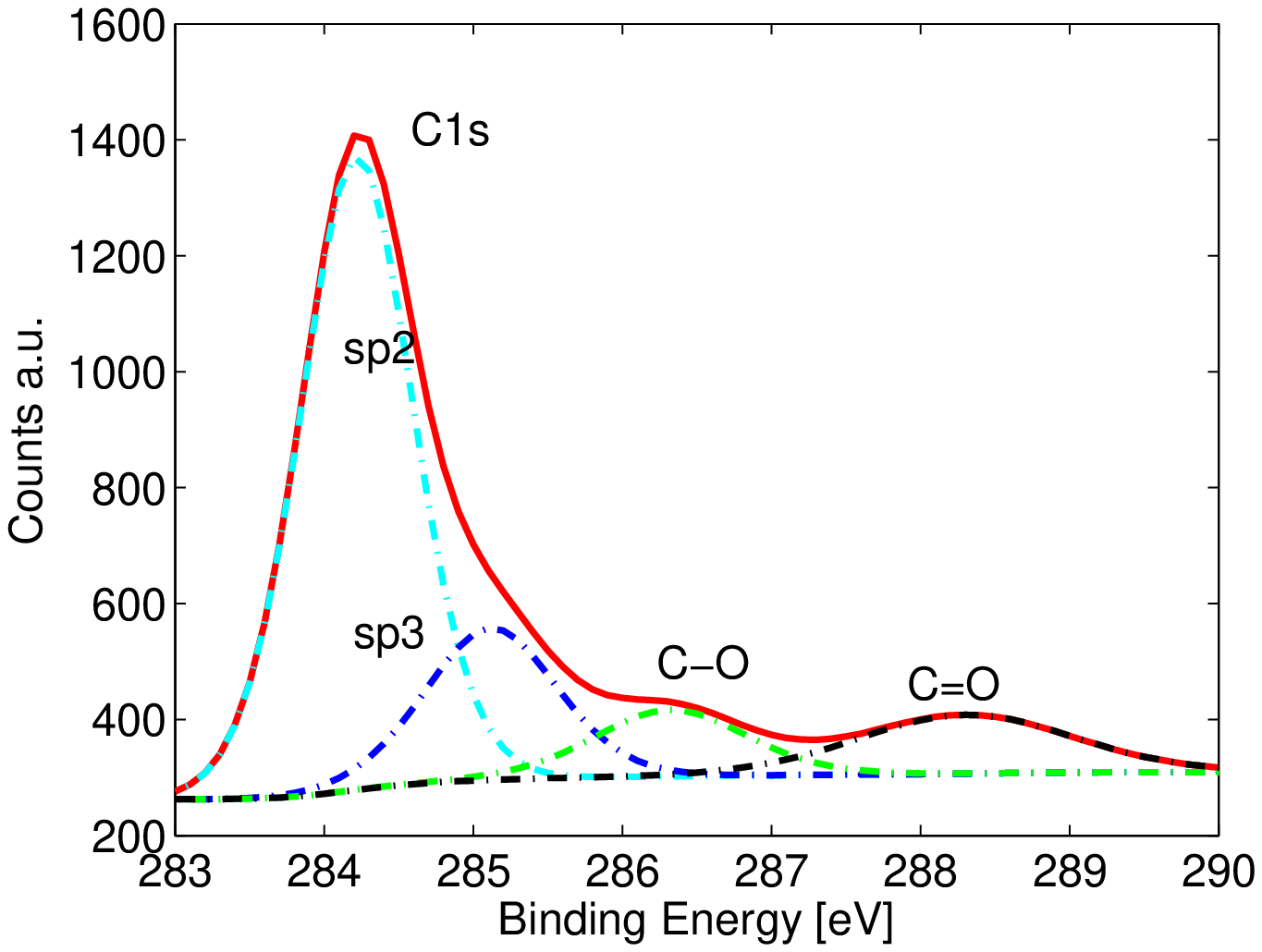}
                \caption{}
                \label{fig:8b}
        \end{subfigure}%
        \caption{(a) Raman spectra for the reference and exposed graphene samples on Cu substrate with naturally accumulated hydrocarbon contamination. (b) XPS results for the graphene sample exposed to oxygen}\label{fig:figure8}
\end{figure*}

Fig.~\ref{fig:8a} shows the Raman spectra of the reference~(unexposed) and exposed graphene samples with naturally accumulated hydrocarbon contamination. It is noted that all the exposed samples show a clear D peak with approximately the same intensity.  The G peak, however, is broadened due to a more disordered carbon network after exposure. Analysis is complicated by the fact that all samples, except the one exposed to oxygen, show an increase in hydrocarbon carbon~(see Tab.~\ref{tab:table2}), which contributes to, not only the broadening, but also the increased intensity of the G peak.  In Tab.~\ref{tab:table2}, it is also apparent that the sp$^2$ atomic concentration is greatly reduced after exposure, from 53.3~at.\% to 34.0~at.\%, indicating that sp$^2$ bonds are being broken and sp$^3$ bonds are being formed. The C-O and C=O peaks are obviously more pronounced in the exposed samples shown in the XPS results in Fig.~\ref{fig:8b} due to EUV induced oxidation of the covering hydrocarbon carbon layer. 

\begin{table*}
\caption{\label{tab:table2}Atomic concentration and total carbon thickness of the reference and exposed graphene samples with naturally accumulated hydrocarbon contamination. The error margin of the data is $\pm1~at.\%$.} 
\begin{ruledtabular}
\begin{tabular}{lccccc}
Sample&Reference&Exposed to background gas&Exposed to H$_2$O&Exposed to H$_2$&Exposed to O$_2$\\
 \hline
C $sp^2$~(at.\%)&53.3&34.0&36.3&36.2&35.7 \\
C $sp^3$~(at.\%)&10.0&13.0&10.6&11.0&10.1 \\
C-O~(at.\%)&1.9&5.0&5.4&7.0&5.1\\
C=O~(at.\%)&2.0&11.2&8.6&6.7&6.4\\
O 1s~(at.\%)&13.0&22.8&23.2&22.8&23.9\\
Cu2p3~(at.\%)&19.8&14.0&15.9&16.3&18.9\\
Total thickness~(nm)&0.98&1.24&1.10&1.06&0.92\\
\end{tabular}
\end{ruledtabular}
\end{table*}

These results show that direct contact of the EUV-induced plasma and the graphene is not a necessary requirement for damage to graphene. The \textit{I(D)/I(G)} ratio~(0.2) found in Fig.~\ref{fig:figure8} is much lower compared with the \textit{I(D)/I(G)} ratio~(1.9) found in Fig.~\ref{fig:figure1} for the sample exposed to EUV with partial water pressure of 1x10$^{-5}$~mbar, indicating that the hydrocarbon carbon layer is an effective barrier layer. In addition, in Fig.~\ref{fig:8a}, the \textit{I(D)} intensities are almost the same for all exposed samples, indicating that the process is largely independent of the background gases. Considering these facts, the induced defects are predominantly due to photon and/or photoelectron induced bond cleaving, rather than the EUV-induced plasma. The photoelectrons from the Cu substrate may play an important role in defect generation in graphene, since the Cu substrate has a relatively high photoelectron yield~\cite{baglin1998photoelectron}. These photoelectrons may directly attribute in breaking the carbon sp$^2$ bonds. 

It is important to note that the damage to graphene caused by plasma depends very strongly on the concentration and composition of the background gases, while, on the other hand, the damage to graphene caused by photon and photoelectron is only dependent on the photon flux. It is, therefore, impossible to determine the relative dominance of these two mechanisms, without specifying the background conditions, photon flux and even the type of substrate (photoelectron yield). Judging from the data in Fig.~\ref{fig:figure8}, the contribution to \textit{I(D)/(G)} from photon and photoelectrons is about 0.2. Considering the experimental results shown with Fig.~\ref{fig:figure1}~(where the \textit{I(D)/(G)} is about 1.9), we can conclude that, in those experiments, EUV-induced plasma is the main source for the defect generation. This finding is reinforced by the fact that the photoelectron yield for the SiO$_2$ substrate is much less than that from the Cu substrate.

\section{Conclusion}

We have studied the damage mechanism of CVD grown single layer graphene under EUV irradiation. We found that the residual water in the vacuum chamber, EUV photons, and/or photoelectrons, all contribute to defect generation in graphene during EUV exposure. The experimental data demonstrate that, under EUV radiation in the presence of water, defects were generated through oxidation, resulting in the graphene being etched, forming cracks and holes, which is similar to thermal oxidation of graphene. The oxidation process originates from the dissociation of water by EUV, and the fact that the EUV photons directly break the sp$^2$ bonds forming sp$^3$ bonds, which leads to defects in graphene. The photoelectrons emitted from the substrate can either cause oxidation via dissociating water molecules on the graphene surface, or directly break sp$^2$ bonds, both of which will induce defects in graphene. Our results help understand lifetime considerations for graphene devices in the presence of hard radiation. Furthermore, the EUV-induced oxidation of graphene provides a possible route to resist-free patterning of graphene.

\begin{acknowledgments}
The authors would like to thank Mr. Goran Milinkovic, Mr. Luc Stevens, Mr. John de Kuster, and Dr. Edgar Osorio for the help with sample preparation and experimental measurements. This work is part of the research programme Controlling photon and plasma induced processes at EUV optical surfaces (CP3E) of the Stichting voor Fundamenteel Onderzoek der Materie (FOM) with financial support from the Nederlandse Organisatie voor Wetenschappelijk Onderzoek (NWO).  The CP3E programme is co-financed by Carl Zeiss SMT and ASML, and the AgentschapNL through the EXEPT programme. This research is also carried out under project number M61.3.11430 in the framework of the Research Program of the Materials innovation institute M2i.
\end{acknowledgments}

%

\end{document}